\documentstyle[proceedings,numreferences,epsfig]{tasy}

\begin{opening}
\title{Two analytic continuations of the Lippmann-Schwinger eigenfunctions}
\author{Rafael de la Madrid}
\institute{Departamento de F\'\i sica Te\'orica \\ Facultad de Ciencias
\\ Universidad del Pa\'\i s Vasco \\ 48080 Bilbao, Spain \\
E-mail: wtbdemor@lg.ehu.es}
\end{opening}
\runningtitle{Two analytic continuations} 

\begin{document}

\begin{abstract}
We first present two possible analytic continuations of the Lippmann-Schwinger
eigenfunctions to the second sheet of the Riemann surface, and then we compare
the different Gamow vectors that are obtained through each analytic 
continuation.
\end{abstract}

\section{Introduction}

Gamow vectors should be related with the analytic continuation of the 
Lippmann-Schwinger eigenfunctions to the resonance energy. However, the
analytic continuation of the Lippmann-Schwinger eigenfunctions to the second 
sheet is not unique. In this paper, we address this non-uniqueness by 
constructing and comparing two possible analytic continuations.


\section{Preliminaries}

Before proceeding with their analytic continuations, we recall some of the 
basic properties of the Lippmann-Schwinger eigenfunctions. Let us take a 
simple example, such as the spherical shell potential for zero angular 
momentum:
\begin{equation}
           V(r)=\left\{ \begin{array}{ll}
                                0   &0<r<a  \\
                                V_0 &a<r<b  \\
                                0   &b<r<\infty \, .
                  \end{array} 
                 \right. 
	\label{potential}
\end{equation}
For this potential, the Lippmann-Schwinger equation 
\begin{equation}
       |E ^{\pm}\rangle = |E\rangle +
       \frac{1}{E-H_0\pm i\epsilon}V|E^{\pm}\rangle 
       \label{LSeq}
\end{equation}
has the following solutions in the radial position 
representation for zero angular momentum~\cite{NEWTON,TAYLOR}:
\begin{equation}
      \langle r|E^{\pm}\rangle \equiv 
       \chi ^{\pm}(r;E)= \kappa (E) \,  
     \frac{\chi (r;E)}{{\cal J}_{\pm}(E)}\, ,
      \label{pmeigndu}
\end{equation}
where $\kappa (E)$ is a normalization factor
\begin{equation}
   \kappa (E) =
     \sqrt{\frac{1}{\pi} \, \frac{2m/\hbar ^2}{\sqrt{2m/\hbar ^2 \, E\,}\,}\,} 
     \ ,
     \label{kappa}
\end{equation}
$\chi (r;E)$ is the regular solution,
\begin{equation}
      \chi (r;E)=\left\{ \begin{array}{lll}
               \sin (\sqrt{\frac{2m}{\hbar ^2}E\,}\, r) \quad &0<r<a  \\
               {\cal J}_1(E)e^{i \sqrt{\frac{2m}{\hbar ^2}(E-V_0)\,}\, r}
                +{\cal J}_2(E)e^{-i\sqrt{\frac{2m}{\hbar ^2}(E-V_0)\,}\, r}
                 \quad  &a<r<b \\
               {\cal J}_3(E) e^{i\sqrt{\frac{2m}{\hbar ^2}E\,}\, r}
                +{\cal J}_4(E)e^{-i\sqrt{\frac{2m}{\hbar ^2}E\,}\, r}
                 \quad  &b<r<\infty \, ,
               \end{array} 
                 \right. 
             \label{LSchi}
\end{equation}
and ${\cal J}_{\pm}(E)$ are the Jost functions,
\begin{equation}
     {\cal J}_+(E)=-2i{\cal J}_4(E) \, ,  \label{jfuncio+} 
\end{equation}
\begin{equation}
      {\cal J}_-(E)=2i{\cal J}_3(E) \, .  \label{jfuncio-}
\end{equation}
In terms of the Jost functions, the $S$~matrix is given by
\begin{equation}
      S(E)=\frac{{\cal J}_-(E)}{{\cal J}_+(E)} \, .
     \label{SmatrixE}
\end{equation}

The $\pm i\epsilon$ impose certain boundary conditions on the solutions
of Eq.~(\ref{LSeq}). In the position representation, these boundary
conditions determine the asymptotic behavior of the Lippmann-Schwinger
eigenfunctions:
\begin{equation}
      \langle r|E^+\rangle \sim 
          e^{-ikr}- S(E)e^{ikr} \, , \quad {\rm as \ } r\to \infty \, ,
        \label{bcoLS5}
\end{equation}
\begin{equation}
      \langle r|E^-\rangle \sim
         e^{ikr} - S^*(E)e^{-ikr} \, , \quad  {\rm  as \ } r\to \infty \, ,
        \label{bcoLS6}
\end{equation}
where
\begin{equation}
      k=\sqrt{\frac{2m}{\hbar ^2}E \, } >0 
      \label{wavenumber}
\end{equation}
is the wave number. 

In terms of $E$, the boundary conditions imposed by the
$\pm i\epsilon$ are tantamount to obtaining boundary values from above 
($+i\epsilon$) and from below ($-i\epsilon$) the spectrum of $H$. Thus,
the Lippmann-Schwinger eigenfunctions are to be seen as boundary values
from above and from below the cut. This is 
analogous to the calculation of the advanced and retarded Green functions
$G^{\pm}(r,s;E)$. These functions are boundary values on the upper and
lower rims of the cut of the Green function $G(r,s;z)$:
\begin{equation}
      G^{\pm}(r,s;E) = \lim_{{\rm Im}(z) \to 0}G(r,s;z) \, ,
       \qquad z=E\pm i {\rm Im}(z), \ {\rm Im}(z) >0 \, .
       \label{gpmE}
\end{equation}

As is well known, for the potential (\ref{potential}) all these functions
(the Green function, the $S$ matrix, and the Lippmann-Schwinger 
eigenfunctions) depend explicitly not on $E$ but on $k$. This is why it is 
convenient to work with $k$ rather than with $E$. In terms of $k$, the 
Lippmann-Schwinger eigenfunctions (\ref{pmeigndu}) read as
\begin{equation}
     \chi ^{\pm}(r;k)=
     \sqrt{\frac{1}{\pi} \, \frac{2m/\hbar ^2}{k}\,}  \,  
     \frac{\chi (r;k)}{{\cal J}_{\pm}(k)}\, .
      \label{kpmeigndu}
\end{equation}
These eigenfunctions are normalized to $\delta (E-E')$. In order to normalize
them to $\delta (k-k')$, we define
\begin{equation}
     \phi ^{\pm}(r;k) \equiv \sqrt{\frac{\hbar ^2}{2m} 2k\,} \chi ^{\pm}(r;k)
     =\sqrt{\frac{2}{\pi}}  \,  
     \frac{\chi (r;k)}{{\cal J}_{\pm}(k)}\, .
      \label{phpmeigndu}
\end{equation}
It is important to notice that in this equation, as well as in 
Eqs.~(\ref{bcoLS5}) and (\ref{bcoLS6}), $k$ is {\it positive}.

We also can write Eq.~(\ref{gpmE}) in terms of $k$. If we denote the complex 
wave number by $q\equiv k+i{\rm Im}(q)$, we have that
\begin{equation}
      G^{+}(r,s;k) = \lim_{{\rm Im}(q) \to 0}G(r,s;q) \, ,
       \qquad q=k+ i {\rm Im}(q), \  {\rm Im}(q),k >0 \, .
       \label{gpk}
\end{equation}
Thus, $G^+$ is the boundary value of $G$ on the positive real line. In the
same vein, $G^-$ is the boundary value of $G$ on the negative real line:
\begin{equation}
      G^{-}(r,s;-k) = \lim_{{\rm Im}(q) \to 0}G(r,s;q) \, ,
       \qquad q=-k+ i {\rm Im}(q), \ {\rm Im}(q),k >0 \, .
       \label{gmk}
\end{equation}

To finish this section, we recall that it is possible to obtain one of the 
Jost functions from the other one by way of the following relation:
\begin{equation}
        {\cal J}_+(k)= {\cal J}_-(-k) \, , \quad k>0 \, .
       \label{jpminof}
\end{equation}
It also holds that
\begin{equation}
      - \chi (r;k)= \chi (r;-k) \, , \quad k>0 \, .
      \label{siofchi}
\end{equation}
Equations~(\ref{phpmeigndu}), (\ref{jpminof}) and (\ref{siofchi}) yield
\begin{equation}
      \phi ^+(r;k)= -\phi ^-(r;-k) \, , \quad k>0 \, ;
      \label{siofphi}
\end{equation}
that is, we can obtain one of the Lippmann-Schwinger eigenfunctions from the 
other one.

\section{Two possible analytic continuations}

After recalling some basic results in the previous section, we are 
in a position to analytically continue the Lippmann-Schwinger eigenfunctions 
into the second sheet of the Riemann surface. We shall examine
two possible analytic continuations, which will be referred to as
Continuation~1 and Continuation~2.

\subsection{Continuation~1}


By means of Eq.~(\ref{siofphi}), we can obtain one of the Lippmann-Schwinger 
eigenfunctions in terms of the other one. What is more, in analogy
with Eqs.~(\ref{gpk}) and (\ref{gmk}), the Lippmann-Schwinger eigenfunctions 
should be seen as limiting values on the positive and on the negative real 
$k$-axis of one and the same function, which we shall denote by $\phi _1$. In 
order to construct $\phi _1$, let us define the Jost function ${\cal J}_1(q)$ 
in such a way that ${\cal J}_1(q)$ coincides with ${\cal J}_+(k)$ on the 
positive $k$-axis and with ${\cal J}_-(k)$ on the negative 
$k$-axis.\footnote{Obviously, ${\cal J}_1(q)$ is just ${\cal J}_+(q)$, although
it is better to use the subindexes $+/-$ only to refer to boundary
values.} The eigenfunction $\phi _1$ is defined as 
follows:\footnote{The subindex 1 refers to Continuation~1.}
\begin{equation}
      \phi _1 (r;q):= \sqrt{\frac{2}{\pi}} \, 
            \frac{\chi (r;q)}{{\cal J}_{1}(q)}  \, .
      \label{phpmeigndu1}
\end{equation}
Then,
\begin{equation}
     \phi _1^+(r;k)=\lim_{{\rm Im}(q) \to 0} \phi _1 (r;q) \, , \quad
      q=k+i{\rm Im}(q) \, , \ k>0 \, ,
\end{equation}
and
\begin{equation}
     \phi _1^-(r;-k)=\lim_{{\rm Im}(q) \to 0} -\phi _1 (r;q) \, , \quad
      q=-k+i{\rm Im}(q) \, , \ k>0 \, .
\end{equation}
That is, we have obtained the Lippmann-Schwinger eigenfunctions as boundary 
values of a unique eigenfunction $\phi _1 (r;q)$ that is defined for all 
complex $q$. The function $\phi _1^+(r;k)$ is the boundary value on the 
positive $k$-axis, whereas $\phi _1^-(r;-k)$ is the boundary value on 
the negative $k$-axis. 

Now, to reach the resonance poles, we continue analytically
$\phi _1^+(r;k)$ into the fourth quadrant and $\phi _1^-(r;k)$ into the
third quadrant. Clearly, this looks like the natural analytic 
continuation.  However, 
resonances seem to need a different analytic continuation, as explained in 
the next subsection.

\subsection{Continuation~2}


The second analytic continuation consists of continuing $\chi ^+(r;E)$
not to the lower half-plane of the second sheet as in Continuation~1, but
to the upper half-plane of the second sheet. In terms of $k$, this means that
instead of continuing $\phi ^+(r;k)$ into the fourth quadrant as 
in Continuation~1, we analytically continue $\phi ^+(r;k)$
into the third quadrant. For the out states, Continuation~2
means that $\chi ^-(r;E)$ is not continued into the upper but into 
the lower half-plane of the second sheet. In terms of $k$, 
$\phi ^-(r;k)$ is not continued into the third but into the fourth 
quadrant.

If we denote the analytic continuation of ${\cal J}_-(k)$ by 
${\cal J}_2(q)$,\footnote{Obviously, ${\cal J}_2(q)$ is just ${\cal J}_-(q)$.} 
we can define the 
eigenfunction\footnote{The subindex 2 refers to Continuation~2.}
\begin{equation}
      \phi _2 (r;q):= \sqrt{\frac{2}{\pi}} \, 
            \frac{\chi (r;q)}{{\cal J}_{2}(q)} 
      \label{phpmeigndu2}
\end{equation}
so that
\begin{equation}
     \phi _2^+(r;-k)=\lim_{{\rm Im}(q) \to 0}  -\phi _2 (r;q) \, , \quad
      q=-k-i{\rm Im}(q) \, , \ k>0 \, ,
\end{equation}
and
\begin{equation}
     \phi _2^-(r;k)=\lim_{{\rm Im}(q) \to 0} \phi _2 (r;q) \, , \quad
      q=k-i{\rm Im}(q) \, , \ k>0 \, .
\end{equation}

Notice that
\begin{eqnarray}
      \phi _2^+(r;-k) = \phi _1^-(r;k) \, , \\
      \phi _2^-(r;k) = \phi _1^-(r;-k)  \, ; 
\end{eqnarray}
that is, when the energy is real or lies on the first sheet, Continuations~1
and 2 yield the same Lippmann-Schwinger eigenfunctions. However, when the
energy lies on the second sheet, Continuations~1 and 2 yield different
Lippmann-Schwinger eigenfunctions.

\section{Differences between Continuations~1 and~2}

Although Continuations~1 and 2 yield the same results for real energies
(thus, the standard results of scattering theory are unchanged), they do
yield different Gamow vectors and different complex 
basis expansions, as we are going to outline in this section.

\subsection{First difference: Gamow vectors}

The analytic continuation of $\phi _1^+(r;k)$ to the fourth quadrant has 
poles. These poles are the same as those of the $S$ matrix. The Gamow vector 
corresponding to the resonance pole $k_n$ is given by
\begin{equation}
   \langle r|k_n ^+ \rangle _1 = A_{n,1} \, 
        {\rm res} \left[ \phi _1^+(r;q) \right]_{q=k_n} \, ,
\end{equation}
where $k_n$ is a zero of ${\cal J}_+(q)\equiv {\cal J}_1(q)$ in the 
fourth quadrant, and $A_{n,1}$ is a normalization factor. The Gamow vector 
corresponding to a zero $-k_n^*$ of ${\cal J}_+(q)$ in the third 
quadrant is given by
\begin{equation}
   \langle r|{\frac{\ \, }{\ \, }k_n^*} ^- \rangle = B_{n,1} \,
        {\rm res} \left[ \phi _1^-(r;q) \right]_{q=-k_n^*} \, ,
\end{equation}
where $B_{n,1}$ is a normalization factor.

By contrast, the continuation of $\phi _2^-(r;k)$ is analytic on the whole
fourth quadrant. According to Continuation~2, the Gamow vector corresponding to
the wave number $k_n$ is given by
\begin{equation}
   \langle r|k_n ^- \rangle _2 = A_{n,2} \, \phi _2^-(r;q=k_n) \, .
\end{equation}
The Gamow vector corresponding to $-k_n^*$ is, according to Continuation~2, 
given by
\begin{equation}
   \langle r|\frac{\ \, }{}{k_n^*}^+ \rangle _2 = 
         B_{n,2} \, \phi _2^+(r;q=-k_n^*) \, .
\end{equation}    
$A_{n,2}$ and $B_{n,2}$ are normalization factors.

For the potential we are considering, $\langle r|k_n ^+ \rangle _1$
and $\langle r|k_n ^- \rangle _2$ are proportional to each other. However,
their wave-number representations are quite different. In fact,
\begin{equation}
   \langle ^+k|k_n ^+ \rangle _1 \equiv \mbox{a somewhat awkward distribution,}
\end{equation}
whereas~\cite{DIS}
\begin{equation}
   \langle ^-k|k_n ^- \rangle _2 \equiv c_n \delta (k-k_n) \equiv
      d_n \frac{1}{k^2-k_n^2} \, ;
\end{equation}
that is, Continuation~2 yields Gamow vectors whose wave-number representations 
are given by the complex delta function and by the Breit-Wigner amplitude 
(up to normalization constants $c_n$ and $d_n$, and up to some technical 
details).

\subsection{Second difference: Complex basis expansions}

Continuation~2 enables the expansion of the wave functions in terms of Gamow 
vectors and a background. This expansion is possible because certain integral 
at infinity vanishes. However, Continuation~1 does not enable this
expansion in a clear way.

\section{Conclusion}

The solutions of the Lippmann-Schwinger equation are to be interpreted as 
boundary values on the upper and lower rims of the cut. Since the upper 
(lower) rim of the cut corresponds to the positive (negative) $k$-axis, 
these boundary values have an easier interpretation in the $k$-plane. 

The Lippmann-Schwinger eigenfunctions can be analytically continued into the
second sheet in two different ways:

--According to Continuation~1, the eigenfunction $\langle r|k^+\rangle _1$ 
(respectively $\langle r|\frac{\ \, }{}k^-\rangle _1$) is analytically 
continued into the fourth (respectively third) quadrant of the complex 
$k$-plane. Continuation~1, although very natural, yields Gamow vectors whose 
wave-number representation is given by an awkward distribution.

--According to Continuation~2, the eigenfunction $\langle r|k^-\rangle _2$ 
(respectively $\langle r|\frac{\ \, }{}{k^*}^+\rangle _2$) is analytically 
continued into the fourth (respectively third) quadrant of the complex 
$k$-plane. Continuation~2 yields Gamow vectors whose wave-number 
representation is given by the complex delta function and by the Breit-Wigner 
amplitude!

Either continuation is mathematically meaningful and has both advantages and 
drawbacks. Continuation~1 has the advantage of providing a very natural 
analytical continuation. Continuation~2 implies an awkward analytic 
continuation, but it provides a link between the Gamow vectors and the 
Breit-Wigner amplitude.  

As of November 21, 2003, it is not known which of these solutions is the right
one, and why the other should not be used. Work on this direction is under
way.

\acknowledgements

This paper grew out of a question the author asked Prof.~Alfonso 
Mondrag\'on. The author wishes to thank Alfonso Mondrag\'on for enlightening,
stimulating conversations on this and other topics. Thanks are also due to
Prof.~Lidia Ferreira for her invitation to participate in the
workshop ``Time Asymmetric Quantum Theory: The Theory of Resonances,'' Lisbon
(2003).

Research supported by the Basque Government through reintegration 
fellowship No.~BCI03.96.



\begin{thebibliography}{99}

\bibitem{NEWTON} R.~G.~Newton, 
{\it Scattering Theory of Waves and Particles}, McGraw-Hill, New York (1966); 
2nd edition, Springer Verlag, New York (1982).

\bibitem{TAYLOR} J.~R.~Taylor, {\it Scattering theory}, Jhon Wiley \& Sons, 
Inc., New York (1972).

\bibitem{DIS} R.~de la Madrid, 
Ph.D.~Thesis, Universidad de 
Valladolid (2001). Available at 
\texttt{http://www.ehu.es/$\sim$wtbdemor/}.


\end{thebibliography}
\end{document}